\documentclass[11pt]{article}
\usepackage{moriond-photos}
\usepackage{epsfig}
\usepackage{floatflt}
\usepackage{xspace}
\usepackage{amsmath}
\usepackage{amssymb}

\def\be{\begin{equation}}
\def\ee{\end{equation}}
\def\bea{\begin{eqnarray}}
\def\eea{\end{eqnarray}}

\newcommand{\tcite}[1]{~$\!$\cite{#1}\xspace}
\newcommand{\pcite}[1]{~$\!$\cite{#1}\xspace}
\newcommand{\as}{\alpha_s}
\newcommand{\hl}{\mathcal{H_L}}
\newcommand{\hr}{\mathcal{H_R}}
\newcommand{\cS}{\mathcal{S}}
\newcommand{\qbar}{{\bar q}}

\begin{document}
\begin{flushright}
  MC--TH--2003--5\\
  LPTHE--P03--09\\
  hep-ph/0305232 
\end{flushright}
\vspace*{-1.3cm} 
\vspace*{4cm} 

\title{Theory and phenomenology of non-global
  logarithms\footnote{Based on talks
    presented at the XXXVIIIth Rencontres de Moriond `QCD and
    high-energy hadronic interactions'.}}

\author{R.~B.~Appleby$^1$ and G.~P.~Salam$^2$}
\address{$^1$Theory Group, Department of Physics and Astronomy, \\
  Schuster Laboratory, University of Manchester, Manchester, M13 9PL,
  England.\smallskip\\
  $^2$ LPTHE, Universit\'e Paris VI and VII and
  CNRS UMR 7589, Paris 75005, France
}

\maketitle\abstracts{
  We discuss the theoretical treatment of
  non-global observables, those quantities that are sensitive only to
  radiation in a restricted region of phase space, and describe how
  large `non-global' logarithms arise when we veto the energy flowing
  into the restricted region. The phenomenological impact of
  non-global logarithms is then discussed, drawing on examples from
  event shapes in DIS and energy-flow observables in 2-jet systems. We
  then describe techniques to reduce the numerical importance of
  non-global logarithms, looking at clustering algorithms in energy
  flow observables and the study of associated distribution of
  multiple observables.}

\section{Introduction and Theory}
\label{sec:intro}

Recently a distinction has been introduced between so-called global
and non-global QCD observables.~\cite{Dasgupta:2001sh} The former are
sensitive to emissions in all directions, while the latter are
sensitive only to emissions in some restricted angular region, for an
example a jet or a hemisphere.  Obvious examples of non-global (NG)
observables are properties of individual jets (invariant mass, numbers
of subjets). Many other common QCD observables are also non-global,
including definitions of diffraction based on rapidity gaps (whether
in terms of particles or energy flow); isolation criteria for photons
(or other particles); distributions of interjet energy flow; or even
the original Sterman-Weinberg jet definition.\pcite{Sterman:1977wj}

The question of globalness becomes of particular relevance whenever
one places a severe restriction on the energy $E$ (or in some cases,
transverse energy) flowing into the observed region. In such a
situation the perturbative series develops logarithmically enhanced
terms at all orders, at the very least single logs $\as^n \ln^n E/Q$,
where $Q$ is the hard scale. For $E\ll Q$ such a series needs to be
resummed. For global observables it has been
shown\tcite{Catani:1992ua} that the resummation can be carried out to
single logarithmic (SL) accuracy, essentially by using the
approximation of independent emission off the underlying Born event.

\begin{floatingfigure}{0.4\textwidth}
  \mbox{}\hspace{-0.04\textwidth}\epsfig{file=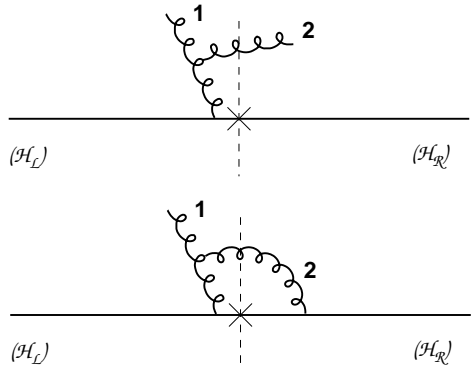,width=0.39\textwidth}
  \caption{Real and virtual contributions to the emission of gluon (2)
    from a $q{\bar q}g$ system.\label{fig:2gloop}}
\end{floatingfigure}
Until recently it had universally been assumed\tcite{Assume} that this
approximation could be used more generally. However it turns out that
for non-global observables, SL resummation is more
subtle.\pcite{Dasgupta:2001sh} This is illustrated in
fig.~\ref{fig:2gloop}, which shows left and right hemispheres ($\hl$,
$\hr$) of a 2-jet $e^+e^-$ event, and two emissions going into
opposite hemispheres such that $Q \gg E_1 \gg E_2$. A global
observable would for example measure the sum of the two gluon
energies. Since $E_1 \gg E_2$, placing a restriction $E_{\max}$ on the
sum is equivalent to placing it directly on $E_1$ and there is
cancellation between the real production and virtual loop
contribution for gluon 2. Because of this cancellation, one is free to
`mistreat' the way gluon $2$ is emitted and pretend it is emitted from
the simpler $q{\bar q}$ system, ignoring the presence of gluon $1$ ---
in other words one can make an independent emission approximation.

Now suppose we have an observable that measures the energy only in
$\hr$. The limit is placed just on gluon $2$, $E_2 < E_{\max}$. On the
other hand the loop contribution has an effective limit
$E_{2,\mathrm{virtual}} \lesssim E_1$ and the mismatch between these
two limits leads to a logarithmic enhancement $\ln E_1/E_2$. After
integrating over $E_1$ one finds an overall contribution $\as^2 \ln^2
Q/E_{\max}$. Making an independent emission approximation, one would
obtain the wrong coefficient for this term. The difference between the
true answer (based on the coherent emission of gluon $2$ from the
$q\qbar g_1$ system) and the independent emission result is termed a
`non-global logarithm' (NGL).

\begin{floatingfigure}{0.4\textwidth}
  \mbox{}\hspace{-0.04\textwidth}\epsfig{file=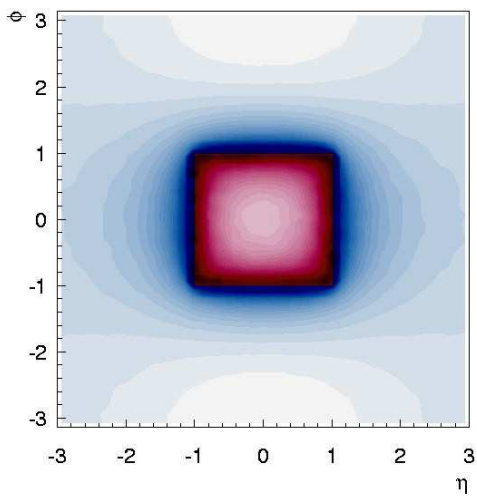,
    width=0.39\textwidth} 
  \caption{Contribution to the second order NGL for measurement in a
    square patch, shown as a function of rapidity and azimuth. Darker
    shades correspond to larger contributions.\label{fig:2gcol}}
\end{floatingfigure}
At this two-gluon level, non-global logarithms are essentially an edge
effect: it is only close to the boundary between measurement and
non-measurement that one is sensitive to the difference between
independent emission and the true two-gluon emission pattern. This is
illustrated in figure~\ref{fig:2gcol} which shows (through the colour
shading) the contribution to the NGL as a function of the two gluons'
rapidities $\eta$ and azimuths $\phi$ for the case in which the
`measurement' is carried out in a square patch $|\eta| < 1$, $|\phi| <
1$.\footnote{When viewed in colour, the blue and red shadings are
  therefore respectively for the unmeasured and measured gluons.} One
sees clearly that the largest contribution to the NG term is
concentrated around the borders of the patch.

In the region where $\as \ln Q/E_{\max}$ is of order $1$, one needs to
understand non-global effects at all orders $\as^n \ln^n Q/E_{\max}$.
This amounts to accounting for the coherent radiation into the
observed region of soft gluons from arbitrarily complex ensembles of
harder (but still energy-ordered) gluons in the non-observed region.
It turns out that given a na{\"\i}ve resummed calculation based on an
independent emission picture, non-global effects can be accounted for
by a multiplicative correction factor $\cS(t)$, where $t$ is the
running-coupling generalisation of $\frac{\as}{2\pi} \ln Q/E_{\max}$:
\begin{equation}
  \label{eq:t}
  t = \int_{E_{\max}}^Q \frac{dE_t}{E_t} \frac{\as(E_t)}{2\pi}\,.
\end{equation}
It is sometimes useful also to write the expansion of $\cS$, $\cS(t) =
\sum_{i=2}^\infty \,\cS_i \,t^i$, and for example the two-gluon
contribution discussed above gives us $\cS_2$. The higher order terms
are more difficult to calculate because of the complicated colour
structures that appear for emission from multi-gluon configurations,
and also because of the geometry. The colour problem has so far only
been partially solved: using the large-$N_c$ approximation one can
restrict one's attention to planar graphs, equivalent to considering
emission from sets of independent colour dipoles.\pcite{BCM}

\begin{figure}
  \begin{minipage}{0.48\textwidth}
    \epsfig{figure=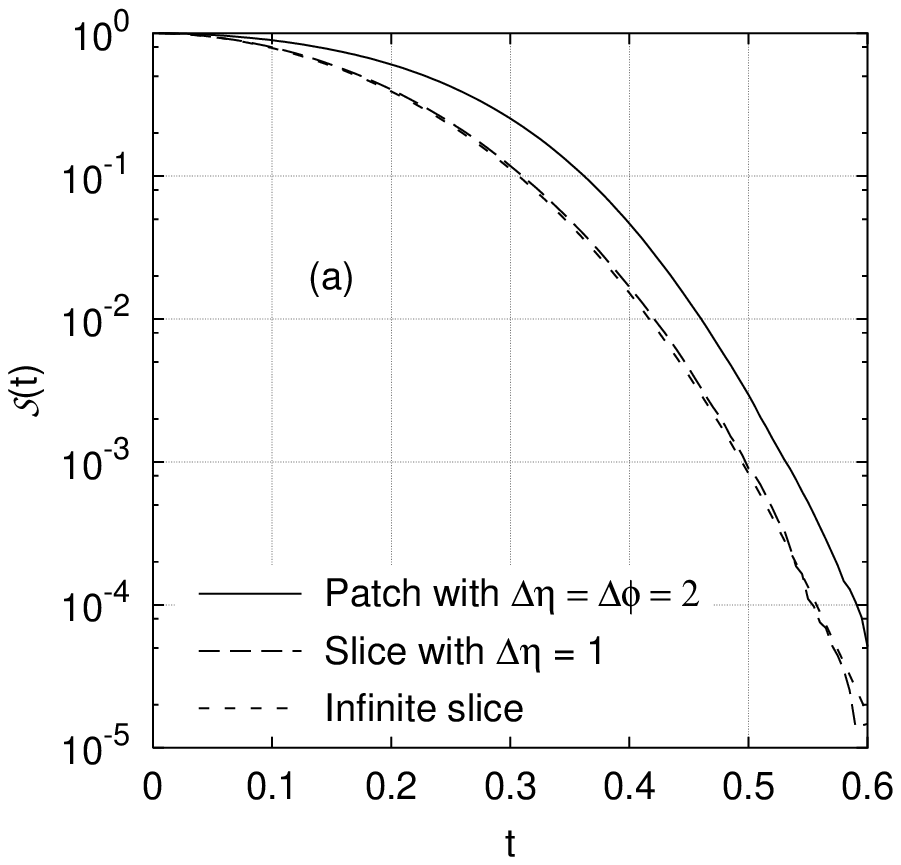,width=\textwidth} 
  \end{minipage}
  \hfill
  \begin{minipage}{0.48\textwidth}\vspace{-5pt}
    \epsfig{figure=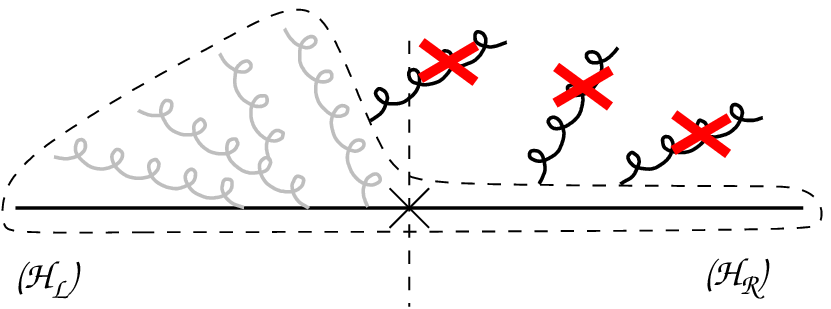,width=0.9\textwidth}\bigskip\medskip\\
    \epsfig{figure=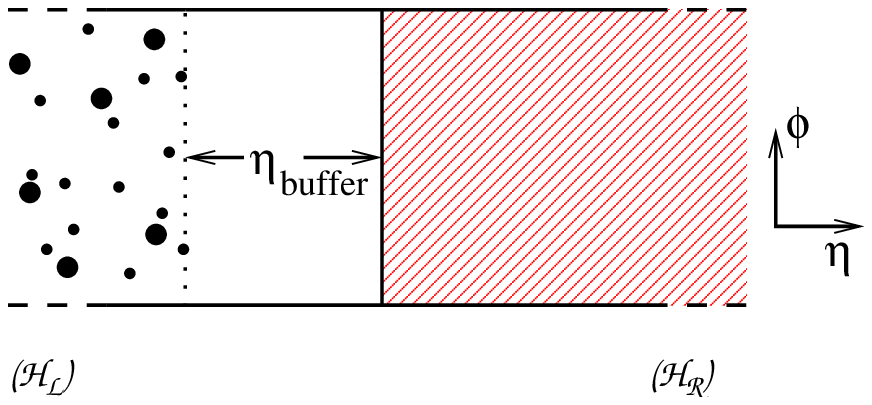,width=\textwidth}%
  \end{minipage}\\
  \begin{minipage}[t]{0.48\textwidth}
    \caption{The function $\cS(t)$ for different geometries of
      observed region in a two-jet event. Slices are defined as
      containing all azimuths but a restricted range of rapidity.
      \label{fig:St}}
  \end{minipage}
  \hfill
  \begin{minipage}[t]{0.48\textwidth}\vspace{-5pt}
    \caption{Above: forbidding coherent radiation from an unobserved
      region (grey gluons) into the observed region (black gluons with
      crosses); below: schematic representation of the actual
      distribution of gluons in $\eta$ and $\phi$ (observed region is
      hashed).
      \label{fig:buffer}}
  \end{minipage}\\
  
\end{figure}

Two equivalent approaches have been proposed to deal with the geometry
dependence. In practice, the simplest way of solving the problem of
the geometry dependence is through a Monte Carlo branching
algorithm,\pcite{Dasgupta:2001sh} in which the original $q \qbar$
dipole branches into two dipoles $qg$ and $g \qbar$. As one increases
the logarithm $t$ each new dipole can itself branch and one
iteratively builds up an ensemble of energy-ordered gluons with the
correct (large-$N_c$) angular distribution.  This algorithm is similar
to that of the Ariadne event generator.\pcite{Ariadne} One then
determines $\cS(t)$ by taking the number of events at scale $t$ that
are free of emissions in the observed region and dividing it by the
number of events that would have been expected on the basis of an
independent emission picture.

Results for $\cS(t)$ are shown in fig.~\ref{fig:St} for various
geometries of observed regions. From a theoretical point of view the
most remarkable feature of these curves is that modulo normalisation,
they all have the same $t$-dependence (in contrast, the fixed order
$\cS_2$ terms differ by more than a factor of two between the
different geometries).  The explanation proposed for this
observation\tcite{Dasgupta:2002dc} was so the so-called `buffer
mechanism', illustrated in fig.~\ref{fig:buffer}: the hypothesis is
that the easiest way of forbidding `secondary' emissions into the
observed region is actually to forbid primary emissions close to the
observed region.  Accordingly for intermediate scales $t' < t$ one
finds a buffer region, free of emissions, surrounding the observed
region. Larger differences $t-t'$ imply a larger buffer region. As a
result the dynamics governing the large-$t$ behaviour of $\cS$ mostly
occurs far from the observed the region and is not affected by the
geometry of the observed region. Assuming that the buffer region is of
size $\delta \eta_\mathrm{buffer} \simeq c C_A (t - t')$ one comes to
the conclusion that
\begin{equation}
  \label{eq:asympS}
  \cS(t) \sim e^{-2 c C_A^2 t^2} .
\end{equation}
These arguments were placed on a mathematically sound footing by
Banfi, Marchesini and Smye\tcite{Banfi:2002hw}. Firstly they
introduced an alternative but equivalent approach to the problem of
the geometry, in terms of a non-linear integral equation:
\begin{equation}
  \label{eq:BMS}
          \partial_t \,\cS_{ab}(t) = \int_{\mathrm{unmeasured}}
        \frac{d^2\Omega_k}{4\pi}\, w_{ab}(k) \left(U_{abk}^{(0)}(t)\,
          \cS_{ak}(t)\, 
          \cS_{kb}(t) - \cS_{ab}(t)\right),
\end{equation}
where $\cS_{ab}(t)$ is the non-global correction factor for a dipole
$ab$, $w_{ab}(k)$ is the weight for emission of a soft gluon $k$ from
$ab$ and $U^{(0)}_{abk}(t)$ accounts for the different `primary'
emission contributions for dipoles $ak,kb$ compared to dipole $ab$.
They were then able to demonstrate the existence of scaling solutions
to \eqref{eq:BMS}, proving the buffer mechanism, with $c$ evaluated
numerically to be $2.5\pm0.25\%$. They also evaluated a number of
subleading corrections. They pointed out however that in practice
\eqref{eq:asympS} applies only for $t \gtrsim 0.5$, i.e.\ very
asymptotic values of $t$. Phenomenologically, $t$ is limited to be
$\lesssim 0.2$ and one should use the full solutions to $\cS$.

\subsection{Warnings for practitioners, a.k.a.\ Zoology}

The definition of a non-global observable, given above, is one that is
sensitive only to emissions (from the Born event) in a restricted
angular region. However the situations in which NGLs can appear are
rather subtle.  

Firstly there exist observables that measure only a subset of the
particles, but which nevertheless are global. An example is the
heavy-hemisphere invariant squared mass in $e^+e^-$. Though only one
hemisphere is measured, the observable is global because it is always
the heavier hemisphere that is measured --- in a situation such as
fig.~\ref{fig:2gloop} the heavier hemisphere is always the one with
the harder gluon (1) and one therefore never has the situation of a
harder unobserved particle radiating into the observed region. Other
examples include certain DIS event shapes (e.g. $B_{zE}$, $\tau_{zQ}$)
where the measurement is carried out in one current hemisphere, but
for which conservation of momentum causes an indirect sensitivity to
emissions in the unobserved hemisphere. Such observables are known as
indirectly global.

Another subtle case is that of observables referred to as
\emph{discontinuously} global. Such observables have different
\emph{parametric} sensitivities to emissions in different directions,
e.g.\ $v \sim E_t^2/Q^2$ for emissions in one hemisphere, $v \sim
E_t/Q$ for the other. Placing a limit on $v$ corresponds to different
limits on $E_t$ in the two hemisphere (e.g.\ $\sqrt{v}Q$ and $vQ$
respectively) and one finds NGLs as for simple non-global observables,
but with the appropriate replacement of the integration limits for $t$
in eq.~\eqref{eq:t}.\pcite{Dasgupta:2002dc,Dokshitzer:2003uw}

The most subtle situation is perhaps that of \emph{dynamically}
discontinuously global observables. These are observables which for a
single emission appear global (typically indirectly global). However
they involve non-linearities such that for the configurations of
emissions that are most common (e.g.\ given a certain value of the
observable) they develop different effective parametric dependence on
emissions in different regions.\pcite{Dasgupta:2002dc} One example of
such an observable is the broadening $B_{zE}$ in DIS.

\section{Phenomenological implications}

In this section we will examine the phenomenological impact of NGLs in
a variety of non-global observables. We will look at the
invariant-squared jet mass in DIS\tcite{Dasgupta:2002dc} (although the
conclusions we draw will apply to $e^+e^-$ non-global event shapes as
well) and at 2-jet energy flows.\pcite{Dasgupta:2002bw} In both kinds
of observable we will see that by neglecting the non-global logarithm
suppression factor, one overestimates the distributions and cross
sections by a considerable, and certainly phenomenologically relevant,
amount.

\subsection{NGLs in DIS and $e^+e^-$ event shapes}

Event shapes (ES), and their distributions,\pcite{Dasgupta:2002dc} are
widely used to compare the predictions of perturbative QCD with
experimental observation. As an example of a non-global event shape,
we will use the DIS invariant squared jet mass $\rho$, defined as
\begin{equation}
\rho=\frac{\left(\Sigma_{\mathcal{H}_C}
    P_i\right)^2}{4\left(\Sigma_{\mathcal{H}_C} |\vec{P}_i|\right)^2},
\end{equation}
where the summation is over all particles in the current-hemisphere
($\mathcal{H}_C$).  When calculating the cross section for $\rho$ to
be smaller than some given value, one finds that in the exclusive
limit, small $\rho$, each power of the coupling is accompanied by up
to two powers of $\ln \rho$, associated with with soft and collinear
divergences. These terms need to be resummed. The standard accuracy is
next-to-leading logarithmic, which in this case implies the inclusion
of all single-logarithmic terms, i.e.\ the same accuracy as NG
contributions.

In figure \ref{fig:rho} we show the effect of the non-global
logarithms on the resummed distribution for~$\rho$. The broken line is
the resummation without NGLs, and the solid line shows the result once
they are included through the function $\mathcal{S}$ discussed above.
Neglecting the non-global logarithms leads to an overestimation of the
peak-height by around 40\%, while the effect is smaller in other parts
of the distribution. In practice, the resummed calculation is usually
matched to a fixed-order calculation, which includes the NGLs to order
$\as^2$ and so reduces the impact of neglecting the NGLs in the final
result.

\subsection{NGLs in 2 jet energy flow observables}

Let us now examine the impact of non-global logarithms on energy flow
observables.\pcite{Dasgupta:2002bw} Such observables have attracted
considerable interest in recent time as an infrared-safe way of
studying gaps-between-jets processes\tcite{OdeSter} and the underlying
event in hadron-hadron collisions.\pcite{MW} We shall take as our
observable the total amount of transverse energy flowing into a
restricted region of phase space $\Omega$. The probability for the
amount of transverse energy to be smaller than some value $Q_\Omega$
is
\begin{equation}
\Sigma_{\Omega}(Q_\Omega) =\frac{1}{\sigma}\int_0^{Q_{\Omega}} d\,E_t
\frac{d\sigma}{dE_t}\,.
\end{equation}
As discussed in section~\ref{sec:intro} this can be written as the
product of two contributions,
\begin{equation}
\Sigma_{\Omega}(Q_\Omega) = \mathcal{S}(t)\times \Sigma_P(t)\,,
\end{equation}
\begin{figure}
  \begin{minipage}{0.48\textwidth}\mbox{}\bigskip\medskip\\
    \epsfig{figure=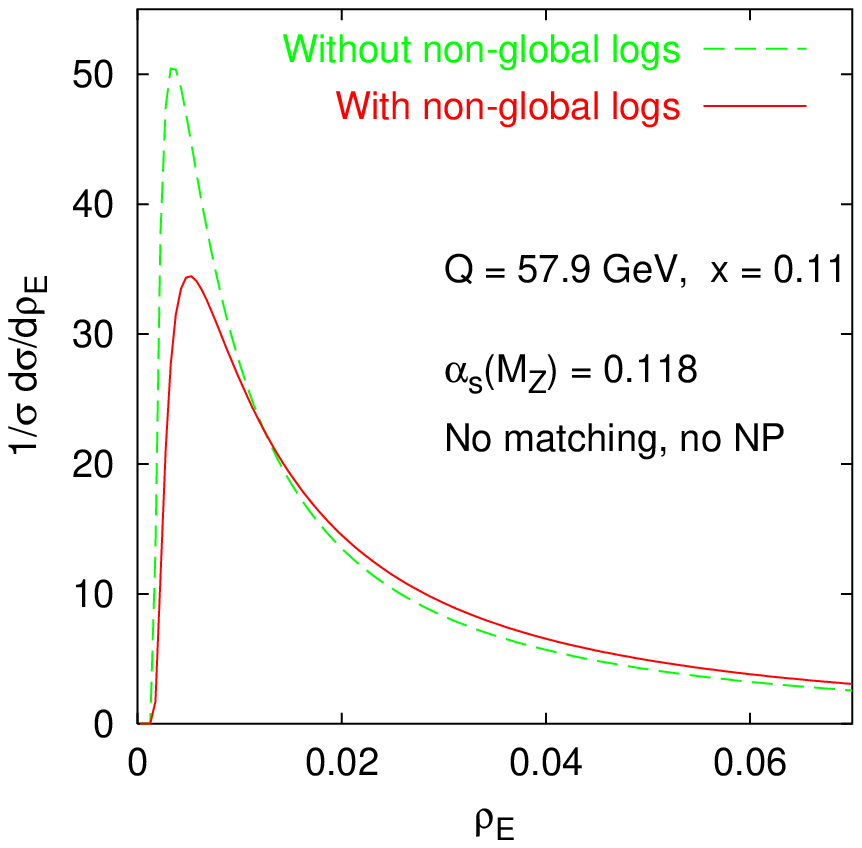,
      width=\textwidth}
    \caption{The impact of non-global logarithms (which enter at the
      NLL level) on a non-global DIS event shape.
      \label{fig:rho}}
  \end{minipage}
  \hfill
  \begin{minipage}{0.48\textwidth}\vspace{-5pt}
    \epsfig{figure=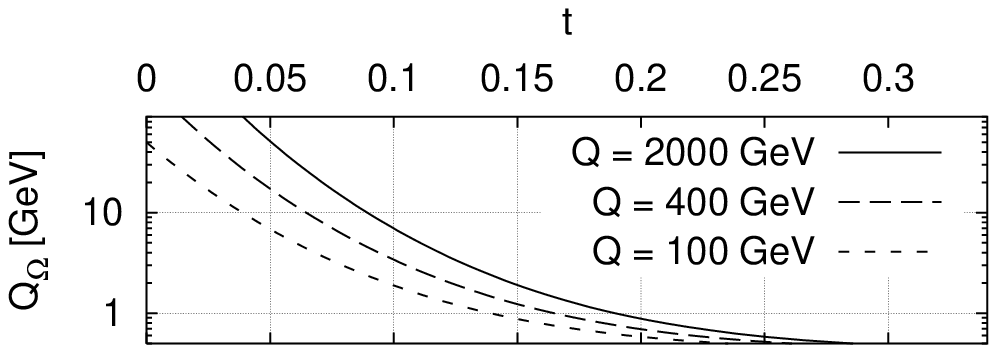,width=\textwidth}\\%
    \epsfig{figure=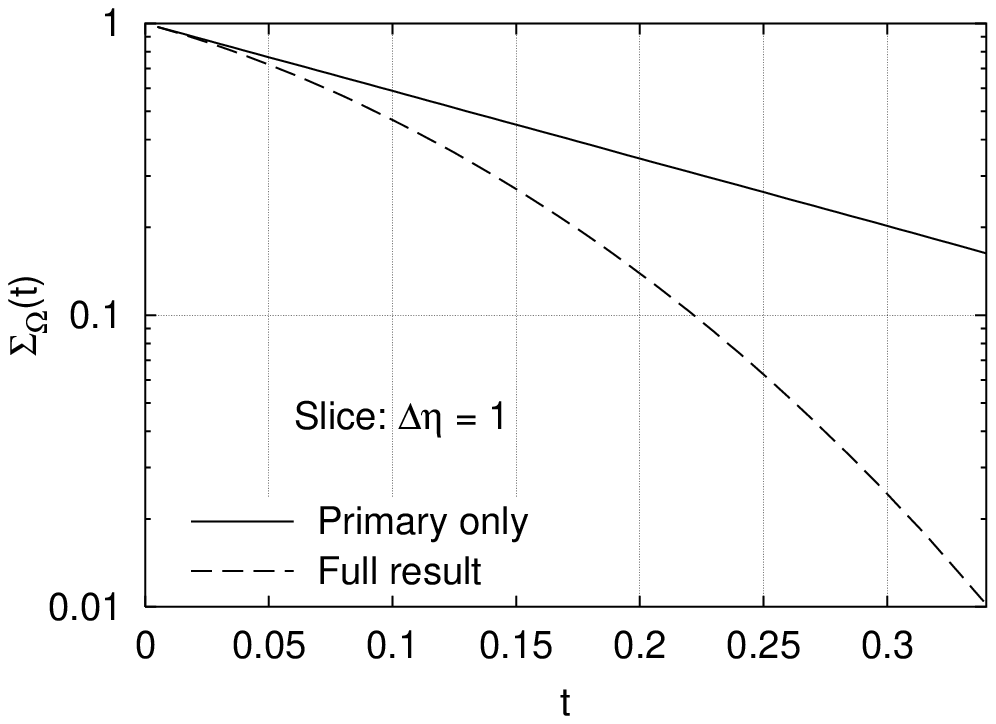,width=\textwidth}%
    \vspace{-6pt} 
    \caption{The impact of non-global logarithms on the 2-jet energy
      flow observable $\Sigma_{\Omega}$. The upper plot shows the
      relation between $Q_\Omega$, $Q$ and $t$.
      \label{fig:slice}}
  \end{minipage}
\end{figure}%
where the function $\Sigma_P$ is based on an independent gluon
emission approximation (`primary' radiation into $\Omega$), while
$\mathcal{S}(t)$ accounts for secondary coherently radiated emission
into $\Omega$ and $t$ is defined as the integral of $\as/2\pi$ between
$Q_\Omega$ and $Q$.
We can understand the phenomenological implications of the non-global
logarithms on this observable by comparing the result for
$\Sigma_{\Omega}$, calculated with only primary emissions, with the
full result for $\Sigma_{\Omega}$, which accounts also for non-global
logarithms. We do this in figure \ref{fig:slice}, where $\Omega$ is a
slice in rapidity of width $\Delta\eta=1.0$.  The plot clearly show
the phenomenological impact of the NGLs on this observable; the
suppression is significant, particularly at larger values of the
effective logarithm $t$. The typical energies of current colliders
corresponds to about $t=0.15$ and if we take this as our reference
value, the inclusion of non-global logarithms increases the
suppression, relative to the primary-only case, by a factor of about
1.65.
Similar results are found for other definitions of $\Omega$, for
example a patch in phase space bounded in rapidity and azimuthal
angle.

\section{Controlling non-global logarithms}

In this section we will describe ways of controlling, or taming,
non-global logarithms. Such a study is important given that our
methods for resumming NG logarithms are both approximate (large-$N_C$
limit) and cumbersome (numerical or asymptotic).
We will look at two different approaches: minimising the numerical
effect of the NGLs by clustering the final state, and also by examining
associated distributions of pairs of observables.

\subsection{Clustering algorithms}

The application of clustering algorithms\tcite{Catani:1993hr} to the
final state is motivated by recent H1 and ZEUS
analyses\tcite{Adloff:2002em} of gaps-between-jets processes at HERA.
In these analyses, the inclusive $k_t$ algorithm is used to define the
hadronic final state, and hence the rapidity gap $\Omega$, and a gap
event is defined by a restriction of the total transverse energy into
$\Omega$.  This observable, sensitive to soft radiation into $\Omega$
only, is clearly non-global, and so sensitive to non-global
logarithmic effects. However, as we will show, the clustering
procedure reduces the numerical importance of the NGLs for this
observable,\pcite{Appleby:2002ke} and we can see why by looking at how
the $k_t$ clustering algorithm works. The essential feature is that in
an iterative algorithm over all the final state particles, it merges
particles of lower transverse momentum into particles with higher
transverse momentum, to produce pseudo-particle or mini-jets. Consider
applying this algorithm to two gluons with strongly ordered transverse
momentum,
\begin{equation}
E_{T,1}\gg E_{T,2},
\end{equation}
where gluon one, directed outside of the region $\Omega$, then in turn
radiates gluon two into the region $\Omega$.  The strong ordering
ensures this configuration produces a non-global logarithm.  When we
cluster this system, gluon two will be clustered into, or merged with,
gluon one (and hence pulled out of the gap) if the two gluons are
sufficiently close in the $(\eta,\phi)$ plane,
\begin{equation}
(\eta_1-\eta_2)^2+(\phi_1-\phi_2)^2<R^2,
\end{equation}
\begin{figure}
  \begin{minipage}{0.47\textwidth}
    \epsfig{figure=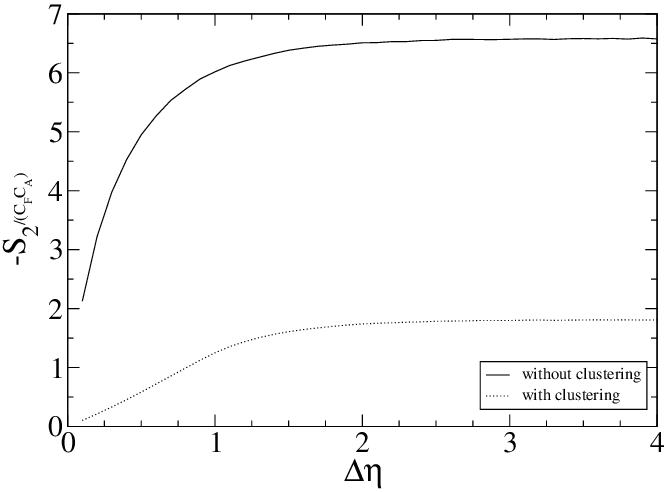,width=\textwidth,height=\textwidth}
    \caption{The leading, $\cS_2$ contribution to the non-global
      logarithm function $\mathcal{S}(t)$, with and without
      clustering.  The saturation of $S_2$ at high $\Delta\eta$ is
      seen for both the non-clustered and the clustered case, with a
      lower saturation value for the latter.
      \label{fig:s2}}
  \end{minipage}
  \hfill
  \begin{minipage}{0.49\textwidth}
    \epsfig{figure=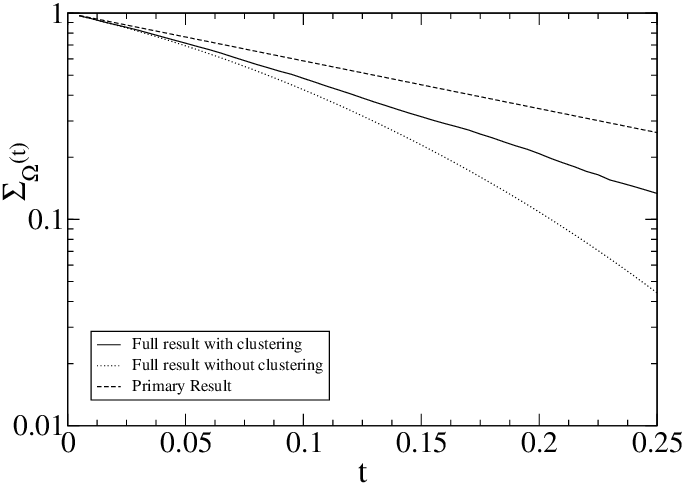,width=\textwidth,height=0.9\textwidth}
    \caption{The phenomenological impact of the $k_t$ clustering
      procedure is to reduce the numerical importance of non-global
      logarithms on this 2-jet energy flow observable.
      \label{fig:sigmaslice}}
  \end{minipage}
\begin{center}
\end{center}
\end{figure}%
where $R$ is a parameter playing the role of a radius in the
algorithm. Therefore to get a kinematical configuration which produces
a NGL, the two gluons need to be sufficiently separated in the
$(\eta,\phi)$ plane to avoid being merged. Hence the $k_t$ algorithm
`cleans up' the restricted region of phase space, $\Omega$, and pulls
soft gluons out of the gap.  Figure \ref{fig:s2} shows the leading,
$\cS_2$ contribution (order $\alpha_s^2$) to the NGL function
$\mathcal{S}(t)$, without clustering (solid line) and with clustering
(broken line.) In both cases $\cS_2$ rapidly saturates at high
$\Delta\eta$, which results from the fact that the NGLs are an edge effect,
and the saturation value for the clustered case is smaller
than that of the non-clustered case. This follows from the fact that,
although the clustering algorithm pulls gluons out of the gap, gluons
can still be sufficiently separated in the $(\eta,\phi)$ plane to
survive clustering and give a significant NGL contribution.
\begin{figure}
\begin{center}
\end{center}
\end{figure}
Figure \ref{fig:sigmaslice} shows the all-orders calculation of the
full function $\Sigma$, where $\Omega$ is taken to be a slice in
rapidity of $\Delta\eta=1.0$.  These figures allow us to see the
phenomenological impact of the non-global logarithms on these
observables, as they show the function $\Sigma_P$ (only primary
logarithms) and the full function $\Sigma$ (primary and non-global
logarithms) with and without clustering. We recall that $t$ is around
$0.15$ to $0.2$ at the energies of current colliders, and as before we
will take $t=0.15$ as our reference value.  The plot shows that the
non-global logarithms cause a considerable suppression of $\Sigma$,
relative to the primary-only result, and that by clustering the final
state this suppression is reduced. At $t=0.15$ the full result without
clustering is suppressed relative to the primary only result by
$1.65$, and this suppression is reduced with clustering to around
$1.2$.

\subsection{Event shape/Energy flow Correlations}

Another way of reducing the phenomenological impact of NGLs that has
been proposed,\pcite{Berger:2003iw} is the study of associated
distributions in two variables. In this work, one combines measurement
of a jet shape $V$ in the whole of phase space (for example thrust,
$V=1-T$) and that of the transverse away-from-jets energy flow
$E_{\mathrm{out}}$.  The former is a global measurement and the latter
is a non-global measurement. If the observable $V$ selects 2-jet-like
configurations, one measures the associated distribution,
\begin{equation}
\Sigma_{\mathrm{2ng}}(Q,V,E_{out}),
\end{equation}
where $Q$ is the hard scale. It has been shown that this distribution
factorizes,\pcite{Dokshitzer:2003uw}
\begin{equation}
  \Sigma_{\mathrm{2ng}}(Q,V,E_{out})=\Sigma(Q,V)
  \cdot\Sigma_{\mathrm{out}}(VQ,E_{\mathrm{out}}),
\end{equation}
where $\Sigma(Q,V)$ is the standard global distribution of $V$ and
$\Sigma_{\mathrm{out}}(VQ,E_{\mathrm{out}})$ contains the logarithmic
distribution in $E_{\mathrm{out}}$. This latter distribution,
containing non-global logarithms is evaluated at the reduced scale
$VQ$, and hence the logarithmic terms will be $(\alpha_s
\log(VQ/E_{\mathrm{out}}))^n$. The work of Berger, K\'ucs and
Sterman\pcite{Berger:2003iw} considered the region in which $VQ$ and $E_{\mathrm{out}}$ 
were comparable, so that the NGLs give a negligible
contribution. Thus, for a restricted subset of appropriately selected
events, it is possible, to `tune out' the non-global logarithmically
enhanced terms in associated distributions.

\section{Conclusion}

To summarise, non-global logarithms are recently discovered
contributions that arise in the distributions of any QCD observable
sensitive only to emissions in a restricted part of phase space.  They
are phenomenologically important and significant progress has been
made in resumming them to all orders in the large-$N_c$ limit.  One of
the main directions of current work focuses on understanding ways of
designing observables so as to reduce the impact of non-global
contributions.

\section*{Acknowledgements}
We are grateful to the organisers and secretarial staff of the QCD
session of Moriond 2003 for a stimulating and enjoyable conference.
RBA would like to acknowledge the University of Manchester for
financial support.


\section*{References}
\begin{thebibliography}{99}

\bibitem{Dasgupta:2001sh}
M.~Dasgupta and G.~P.~Salam,
Phys.\ Lett.\ B {\bf 512} (2001) 323.

\bibitem{Sterman:1977wj}
G.~Sterman and S.~Weinberg,
Phys.\ Rev.\ Lett.\  {\bf 39} (1977) 1436.

\bibitem{Catani:1992ua}
S.~Catani, L.~Trentadue, G.~Turnock and B.~R.~Webber,
Nucl.\ Phys.\ B {\bf 407} (1993) 3.

\bibitem{Assume} For example in references 6, 10, 11 and 12 of
  M.~Dasgupta and G.~P.~Salam,
Acta Phys.\ Polon.\ B {\bf 33} (2002) 3311.


\bibitem{BCM} A.~Bassetto, M.~Ciafaloni and G.~Marchesini,
Phys.\ Rept.\ {\bf 100} (1983) 201.



\bibitem{Ariadne}
L.~L\"onnblad,
Comp.\ Phys.\ Comm.\ {\bf 71} (1992) 15.

\bibitem{Dasgupta:2002dc} M.~Dasgupta and G.~P.~Salam,
JHEP {\bf 0208} (2002) 032.

\bibitem{Banfi:2002hw}
A.~Banfi, G.~Marchesini and G.~Smye,
JHEP {\bf 0208} (2002) 006.

\bibitem{Dokshitzer:2003uw}
Yu.~L.~Dokshitzer and G.~Marchesini,
JHEP {\bf 0303} (2003) 040.

\bibitem{Dasgupta:2002bw}
M.~Dasgupta and G.~P.~Salam,
JHEP {\bf 0203} (2002) 017.


\bibitem{OdeSter}
G.~Oderda and G.~Sterman,
Phys.\ Rev.\ Lett.\ {\bf 81} (1998) 3591.

\bibitem{MW}
G.~Marchesini and B.R.~Webber,
Phys.\ Rev.\ D {\bf 38} (1988) 3419.



\bibitem{Catani:1993hr}
S.~Catani, Yu.~L.~Dokshitzer, M.~H.~Seymour and B.~R.~Webber,
Nucl.\ Phys.\ B {\bf 406} (1993) 187.

\bibitem{Adloff:2002em}
C.~Adloff {\it et al.}  [H1 Collaboration],
Eur.\ Phys.\ J.\ C {\bf 24} (2002) 517.



\bibitem{Appleby:2002ke}
R.~B.~Appleby and M.~H.~Seymour,
JHEP {\bf 0212} (2002) 063.


\bibitem{Berger:2003iw}
C.~F.~Berger, T.~K\'ucs and G.~Sterman,
hep-ph/0303051.





\end{thebibliography}
\end{document}